\documentstyle[preprint,prc,aps]{revtex}

\begin{document}
\draft

\title{ $\pi NN$ and $\pi N\Delta$ formfactors determined from a
microscopic model for $\pi N$ scattering }

\author{ C. Sch\"utz and K. Holinde}

\address{Institut f\"{u}r Kernphysik, Forschungszentrum J\"{u}lich
GmbH, D--52425 J\"{u}lich, Germany}

\maketitle

\begin{abstract}
We determine the $\pi NN$ and $\pi N\Delta$ formfactors from the
$P_{11}$ resp. $P_{33}$ partial wave of $\pi N$ scattering by dressing
corresponding bare vertices with the help of $\pi N$ non--pole
contributions.  The underlying model is based on meson exchange, and
involves nucleon and delta--isobar pole and crossed--pole terms together
with correlated $\pi\pi$--exchange in the $J^P=0^+$ ($\sigma$) and $1^-$
($\rho$) channel.  The results are very similar for $\pi NN$ and $\pi
N\Delta$ and can be roughly parametrized by a monopole with cutoff mass
\rlap{\lower 4pt \hbox{\hskip 1pt $\sim$}}\raise 1pt \hbox{$>$} 500 MeV,
with some variation due to model dependencies. Thus the formfactors are
much less soft than derived before for the $\pi NN$ case by Saito and
Afnan using the same procedure but different $\pi N$ interaction models.
\end{abstract}
\pacs{}

The strong $\pi NN$ (and $\pi N\Delta$) vertex plays an important role
everywhere in nuclear physics. Therefore a precise knowledge of
corresponding formfactors at these vertices is essential in order to
reach a combined and consistent understanding of nuclear phenomena.

Recently Saito and Afnan \cite{Saito94} determined the $\pi NN$
formfactor from the $P_{11}$ partial wave of $\pi N$ scattering. Within
their $\pi N$ interaction model a bare $\pi NN$ formfactor in the
($s$--channel) nucleon pole term gets dressed by phenomenological
separable non-pole terms, through the iteration in a Lippmann--Schwinger
equation. Their result for the dressed formfactor is extremely soft,
corresponding to a monopole cutoff mass $\Lambda_{\pi NN}$ much less
than 400 MeV, which is one reason why their resulting three--body force
contribution to the triton binding energy turns out to be extremely
small ($\simeq-2$ keV).  Moreover, starting from completely different
bare formfactors (monopole masses of 1822 and 323 MeV, respectively),
the authors of Ref. \cite{Saito94} were able to show that within their
model the corresponding dressed formfactors had quite similar, extremely
soft behavior indicating that the requirement to fit the experimental
data puts considerable constraints on the dressed formfactor.

On the other hand there is numerous information \cite{Coon90}, also from
QCD lattice calculations \cite{Liu95}, that the $\pi NN$ formfactor
should be characterized by a monopole cutoff mass around 800 MeV. This
is not nearly as soft as found in Ref. \cite{Saito94} and thus provides
much less suppression of the $\pi NN$ vertex. (Of course it is still
soft compared to the hard $\pi NN$ formfactors used in most $NN$ boson
exchange models; for example, in the full Bonn potential
\cite{Machleidt87}, $\Lambda_{\pi NN}=1.3$ GeV).

Therefore the question arises (which was put already in
Ref. \cite{Saito94}) whether such a determination of the $\pi NN$
formfactor from empirical $P_{11}$ $\pi N$ scattering phase shifts as
performed in Ref. \cite{Saito94} provides indeed an unambiguous,
extremely soft result (being in contrast to other informations
\cite{Coon90,Liu95}) or whether this is a special feature of the $\pi N$
model used in Ref. \cite{Saito94}. It is the purpose of this letter to
address this issue, by starting from an alternative $\pi N$ interaction
\cite{Schuetz94}.

This model developed recently by our group in J\"ulich is based on meson
exchange. It contains, apart from nucleon-- ($N$) and delta--isobar
($\Delta$) s--channel pole terms, non--pole pieces consisting of crossed
N and $\Delta$ exchange and correlated $\pi\pi$ exchange in the $0+$
($\sigma$) and $1-$ ($\rho$) channels as visualized in Fig. 1.  A
satisfactory description of all $\pi N$ scattering phase shifts below
pion production is achieved. It is important to note that the non-pole
pieces which determine the $\pi NN$ formfactor now act with a unified
set of parameters in all partial waves and are therefore strongly tested
by the simultaneous description of all $S$ and $P$ waves.  (This is not
true in Ref. \cite{Saito94} since there only the $P_{11}$ data put a
constraint on the dressed $\pi NN$ formfactor).

In order to derive the renormalized $\pi NN$ formfactor we start from
the 'dressed' vertex function $v^{\pi NN}$
\begin{equation}
v^{\pi NN}\equiv v_0^{\pi NN} + T^{non-pole}_{\pi N}G_{\pi N}\,v_0^{\pi
NN} \;\;\; ,
\label{eq1}
\end{equation}
with the bare vertex $v_0^{\pi NN}$ given by ($p$ being the relative
$\pi N$ momentum)
\begin{eqnarray}
v_0^{\pi NN} (p) &=& \sqrt{3}{f^0_{\pi NN}\over m_\pi \sqrt{4\pi}}
{1\over\sqrt{2\pi}} \; {E_N (p) + \omega_\pi (p) +m_N \over\sqrt{E_N (p)
\omega_\pi (p) (E_N (p) + m_N)}} \;p\; F^{bare} _{\pi NN} (p) \nonumber
\\ &\equiv & {\cal F}^{\pi NN} (p) \; F^{bare} _{\pi NN} (p) \;\; .
\label{eq2}
\end{eqnarray}

Here $F^{bare} _{\pi NN} (p)$ denotes the bare $\pi NN$ formfactor.
(Since the nucleon is a $P$--wave resonance the momentum dependence of
$v_0^{\pi NN}$ is essentially given by the factor of $p$). The $\pi N$
model of Ref. \cite{Schuetz94} is based on time--ordered perturbation
theory; therefore the $\pi N$ propagator $G_{\pi N}$ has to be chosen
accordingly. The non--pole amplitude $ T^{non-pole}_{\pi N}$ is
generated by iterating the non--pole part of the potential,
$V^{non-pole}_{\pi N}$. After partial wave decomposition,
Eq. (\ref{eq1}) reads explicitly
\begin{equation}
v^{\pi NN}(p,Z)=v_0^{\pi NN}(p) + \int_0^\infty q^2dq\,
T^{non-pole}_{\pi N}(p,q;Z) {1\over Z-E_N(q)-\omega_\pi (q) -i \epsilon}
v_0^{\pi NN}(q) \;\;\; ,
\label{eq3}
\end{equation}
$Z$ being the $\pi N$ starting energy.  The physical formfactor $F^{\pi
NN}(p,Z)$ is then obtained from $v^{\pi NN}(p,Z)$ by dividing out the
momentum dependence already inherent in the bare $\pi NN$ vertex.  It is
a function of both $Z$ and $p$ and is normalized to unity at the
physical nucleon pole, i.e. at $Z=m_N$. Consequently
\begin{equation}
F^{\pi NN} (p, Z) = {v^{\pi NN}(p,Z)\over {\cal F}^{\pi NN}(p)} \;
{{\cal F}^{\pi NN}(p_0)\over v^{\pi NN}(p_0, Z=m_N)} \;\; ,
\label{eq4}
\end{equation}
where $p_0$ is the on--shell momentum belonging to $Z=m_N$.

Later we will also present results for the dressed $\pi N\Delta$
formfactor.  All formulas given so far hold correspondingly; only
$v_0^{\pi NN}$ has to be replaced by $v_0^{\pi N\Delta}$ given by
\begin{eqnarray}
v_0^{\pi N\Delta} (p) &=& {1\over\sqrt{6\pi} m_\pi}{f^0_{\pi
N\Delta}\over\sqrt{4\pi}} \sqrt{E_N (p) + m_N\over E_N (p) \omega_\pi
(p)} \; p \; F^{bare} _{\pi N\Delta} (p) \nonumber \\ &\equiv & {\cal
F}^{\pi N\Delta} (p) \; F^{bare} _{\pi N\Delta} (p) \;\; .
\label{eq5}
\end{eqnarray}

In the following we use two different $\pi N$ models 1, 2 (for details,
see Ref. \cite{Schuetz94}), which are based on the same dynamical input
(cf. Fig. 1), but differ in the parametrization of the formfactors in
both the pole and non--pole contributions. The predictions of these
models for the $\pi N$ partial waves of relevance for the calculation of
the $\pi NN$ and the $\pi N\Delta$ formfactor are given in Fig. 2.

Results for the dressed $\pi NN$ formfactor evaluated according to
Eq. (\ref{eq4}), at $Z=m_N$ and as function of p, are shown in Fig. 3
(a), in comparison to a simple monopole parametrization,
$F(p)=(\Lambda^2-m_\pi^2)/(\Lambda^2+p^2)$.
Corresponding results for the $\pi N\Delta$ formfactor are shown in
Fig. 3 (b).  In principle the $\pi N\Delta$ formfactor defined
analogously to Eq. (\ref{eq4}) is a complex quantity since the
delta--isobar lies within the physical range of $\pi N$ scattering. It
turns out however, that the imaginary part of the formfactor is very
small at $Z=m_\Delta$. Therefore we restrict our discussion to its real
part at this place.
For both models the $\pi NN$ and the $\pi N\Delta$ formfactor have very
similar structure.  Deviations occur at smaller momenta, mainly since
the normalization points are different ($p_0 = 227.3$ MeV for $\pi
N\Delta$ and $p_0 = i\, 137.3$ MeV for $\pi NN$).

In Fig. 4, we compare our predictions for the $\pi NN$ formfactor to
corresponding results obtained in Ref. \cite{Saito94}.  (In order to
enable a comparison in the latter case our results have been normalized
to unity in Fig. 4 at $p=0$.)  The results have the following main
features:

(i) For both models 1, 2 the renormalized $\pi NN$ formfactor is much
less soft than found in Ref. \cite{Saito94}, though still softer than
the presently favored monopole with a cutoff mass of 0.8 GeV.

(ii) Results for model 1 and 2 are considerably different although the
non--pole amplitude is generated from identical dynamics and is
constrained by all $\pi N$ partial waves. Obviously the remaining
freedom in parametrizing the (bare) formfactors in the pole terms and
those in the non-pole contributions has substantial effects.  Finally,
we would like to point out that inclusion of further processes in the
non--pole part of the interaction model (like e.g. coupling to the
reaction channel $\pi\Delta$) might also lead to a change of the dressed
$\pi NN$ formfactor.

\bigskip

We thank the authors of Ref. \cite{Saito94} for providing to us the
numerical values of their formfactor results.

\begin{figure}
\caption{Diagrams included in the $\pi N$ potential.}
\label{fig:diags}
\end{figure}

\begin{figure}
\caption{$\pi N$ scattering phase shifts in the $P_{11}$ and the
$P_{33}$ partial wave, as function of the pion laboratory momentum. The
solid (dashed) lines are the results obtained in model 1 (2) of
Ref. \protect\cite{Schuetz94}. The empirical information is taken from
Ref. \protect\cite{Koch80}.}
\label{fig:phases}
\end{figure}

\begin{figure}
\caption{$\pi NN$ (a) and $\pi N\Delta$ formfactor (b) as function of
the square of the pion momentum in the $\pi N$ c.m. system. The solid
(dash--dotted) lines denote the predictions resulting from model 1 (2)
of Ref. \protect\cite{Schuetz94}. The dashed (dotted) lines
represent a conventional monopole formfactor with the cutoff mass
$\Lambda$ = 500 (700) MeV.}
\label{fig:fofa1}
\end{figure}

\begin{figure}
\caption{$\pi NN$ formfactor (normalized to unity for $p=0$) as function
of the pion momentum in the $\pi N$ c.m. system. The solid
(dash--dotted) line denotes our prediction resulting from model 1 (2) of
Ref. \protect\cite{Schuetz94}, whereas the dotted (dashed) line is the
result of model PJ (M1) of Ref. \protect\cite{Saito94}.}
\label{fig:fofa2}
\end{figure}

\end{document}